\documentclass[11pt,reqno]{amsart}
\usepackage[a4paper,margin=25mm]{geometry}
\usepackage{amsmath}
\usepackage{booktabs}
\usepackage{latexsym}
\usepackage{mathrsfs}
\usepackage{amsthm}
\usepackage{amssymb}
\usepackage{amsfonts}
\usepackage{amsbsy}

\usepackage[shortlabels]{enumitem}
\usepackage{url}
\usepackage{array}
\usepackage{pdflscape}
\usepackage{xcolor}
\usepackage{stmaryrd}
\usepackage{verbatim}
\usepackage{braket}
\usepackage{graphicx}
\usepackage{tikz}
\usepackage[hidelinks]{hyperref}
\usepackage{makecell}

\usepackage{setspace}
\newcommand{\Ff}{{\mathbb F}}
\newcommand{\E}{{\mathbb E}}

\newcommand{\Zz}{{\mathbb Z}}

\newcommand{\bv}{{\bf v}}

\newcommand{\cB}{{\mathcal B}}
\newcommand{\cD}{{\mathcal D}}

\newcommand\cc{{\mathcal C}}        %

\newcommand\cF{{\mathbf c}}

\theoremstyle{plain}
\newtheorem{thm}{Theorem}[section]
\newtheorem{lem}[thm]{Lemma}

\theoremstyle{definition}
\newtheorem{defn}{Definition}[section]
\theoremstyle{plain}

\DeclareMathOperator{\ord}{ord}
\DeclareMathOperator{\Span}{Span}

\newcommand{\ceil}[1]{\left\lceil #1\right\rceil}
\newcommand{\floor}[1]{\left\lfloor #1\right\rfloor}

\title{On the maximum length of cyclic MDS codes}
\author{Gaojun Luo}
\address{School of Mathematics, Nanjing University of Aeronautics and Astronautics, Nanjing 211106, China}
\address{Key Laboratory of Mathematical Modelling and High Performance Computing of Air Vehicles, MIIT, Nanjing 211106, China}
\email{gaojun\_luo@nuaa.edu.cn}
\thanks{The research of Gaojun Luo was supported by the National Natural Science Foundation of China under Grant 12401690.}

\author{Xiwang Cao}
\address{School of Mathematics, Nanjing University of Aeronautics and Astronautics, Nanjing 211106, China}
\address{Key Laboratory of Mathematical Modelling and High Performance Computing of Air Vehicles, MIIT, Nanjing 211106, China}
\email{xwcao@nuaa.edu.cn}
\thanks{The research of Xiwang Cao was supported by the National Natural Science Foundation of China under Grant 12571575.}

\author{Shitao Li}
\address{the School of Internet, Anhui University, Hefei, Anhui
	230039, China}
\email{lishitao0216@163.com}

\author{Yang Li}
\address{the School of Physical and Mathematical Sciences, Nanyang
	Technological University, Singapore 637371}
\email{yanglimath@163.com}


\begin{document}
\raggedbottom
\begin{abstract}
We study the maximum length of nontrivial simple-root cyclic maximum distance separable (MDS) codes over $\Ff_q$. We prove $n\le q+1$ if $\gcd(n,q^2-1)>2$, or if the dimension is even and $\gcd(n,q^2-1)>1$. The bound also holds for all lengths whose prime factors lie in $\{2,3,5,7\}$. We obtain further cases involving primes $\ell\ge11$ with $(\ell-1)/2$ prime, under conditions on the exponents and the characteristic. The proofs combine Fourier supports, locality bounds and field descent with arithmetic restrictions on multiplicative orders and prime-adic valuations.
\end{abstract}
\maketitle

\section{Introduction}\label{sec:intro}

Let $q$ be a prime power and let $\Ff_q$ be the finite field with $q$ elements. An $[n,k,d]_q$ \emph{linear code} $\cc$ is a $k$-dimensional subspace of $\Ff_q^n$ with minimum Hamming distance $d$. The Singleton bound \cite{MS} gives $d\leq n-k+1$. The code $\cc$ is \emph{maximum distance separable} (MDS) if $d=n-k+1$. We call such a code an $[n,k]_q$ MDS code. MDS codes with parameters $[n,1]_q$, $[n,n-1]_q$ or $[n,n]_q$ exist for every $n\ge2$. These codes are called \emph{trivial} MDS codes. An MDS code is \emph{nontrivial} if $2\leq k \leq n-2$. 

For fixed $k\ge2$ and $q$, let $m(k,q)$ be the maximum length of a $q$-ary linear MDS code of dimension $k$. Determining this length is a central problem in coding theory. Segre \cite{Segre1955} proposed the MDS conjecture in 1955. The conjecture states that every nontrivial $[n,k]_q$ linear MDS code satisfies $n\le q+1$, except when $q$ is even and $k\in\{3,q-1\}$. In these cases, the conjectured bound is $n\le q+2$. This length problem is equivalent to finding the maximum size of an arc in the projective space $\mathrm{PG}(k-1,q)$; see \cite{BallLavrauw}.

Ball~\cite{Ball} proved the conjecture when $q$ is prime. For $q=p^f$ with $p$ an odd prime and $f>1$, Ball and De Beule~\cite{BallDeBeule} proved the bound $n\le q+1$ for $k\le 2p-2$. The maximum length of MDS elliptic codes has been studied in \cite{chen1994main,Han2023,Han2024,Li2015,Munuera1992}. Deng, Feng, and Vasil'ev~\cite{DFV} studied the lengths of MDS codes with a $2$-transitive permutation automorphism group. They proved that every such $[n,k]_q$ code with $4\le k\le q-3$ satisfies $n\le q+1$. Fan \cite{Fa} proved the conjecture for a range of dimensions when $q$ is odd.

An $[n,k,d]_q$ linear code $\cc$ is \emph{cyclic} if $(c_0,c_1,\cdots,c_{n-1})\in\cc$ implies $(c_{n-1},c_0,\cdots,c_{n-2})\in\cc$. Its permutation automorphism group is transitive on the coordinate positions, but it need not be $2$-transitive. The code $\cc$ can be identified with an ideal of the quotient ring $\Ff_q[x]/(x^n-1)$. A cyclic code of length $n$ over $\Ff_q$ is called \emph{simple-root} if $\gcd(n,q)=1$ and \emph{repeated-root} otherwise. These terms refer to whether $x^n-1$ has distinct or repeated roots in its splitting field.

The existence of $[n,k]_q$ cyclic MDS codes with $n\mid (q-1)$ or $n\mid (q+1)$ has been studied in \cite{Dahl1992,Grassl2015,Li2025,MS,Maruta1998,PedersenDahl,Reed1960}. Let $p$ be the characteristic of $\Ff_q$. The repeated-root case has been settled in \cite{PedersenDahl}. Pedersen and Dahl \cite{PedersenDahl} proved that, for $2\le k\le n-2$, a repeated-root $[n,k]_q$ cyclic MDS code exists if and only if $n=p$. To the best of our knowledge, the maximum length of a nontrivial simple-root cyclic MDS code over $\Ff_q$ with prescribed dimension $k$ remains undetermined in general. Deng, Feng, and Vasil'ev~\cite[Section~5]{DFV} also identify the MDS conjecture for cyclic codes as a problem for further study. All cyclic codes considered below are assumed to be simple-root. 

We study the maximum length of nontrivial cyclic MDS codes. The main results are stated below.

\begin{thm}\label{thm:A}
Let $q$ be a prime power and let $\cc$ be a nontrivial $[n,k]_q$ cyclic MDS code. Suppose that at least one of the following conditions holds:
\begin{enumerate}
	\item $\gcd(n,q^2-1)>2$;
	\item $k$ is even and $\gcd(n,q^2-1)>1$.
\end{enumerate}
Then $n\le q+1$.
\end{thm}

\begin{thm}\label{cor:A}
Let $q$ be a prime power. Every nontrivial $[n,k]_q$ cyclic MDS code satisfies $n\le q+1$ in either of the following cases:
	\begin{enumerate}
		\item $3\mid n$ or $4\mid n$.
		\item There are distinct odd prime divisors $\ell_1,\ell_2$ of $n$ with
		$\gcd(\ell_1-1,\ell_2-1)=2$.
	\end{enumerate}
\end{thm}

\begin{thm}\label{thm:B}
	Let $q=p^f$, where $p$ is a prime and $f\ge1$.
	Let $\cc$ be a nontrivial $[n,k]_q$ cyclic MDS code.
	Let $\ell_1,\ldots,\ell_s$ be distinct primes with $s\ge1$,
	such that $\ell_i\ge11$ and $(\ell_i-1)/2$ is prime
	for every $1\le i\le s$.
	Then $n\le q+1$ in each of the following cases.
	\begin{enumerate}
		\item $n=2^\alpha3^\beta5^\gamma7^\delta$, where
		$\alpha,\beta,\gamma,\delta$ are nonnegative integers.
		
		\item $n=2^\alpha3^\beta5^\gamma7^\delta
		\prod_{i=1}^s\ell_i^{e_i}$, where
		$\alpha\ge1$ and $\beta,\gamma,\delta,e_1,\ldots,e_s$
		are nonnegative integers, such that for each
		$1\le i\le s$, either $e_i\le2$ or
		$\ell_i\ge\lceil p/2\rceil$.
		
		\item $n=\ell_i^e$ for some $1\le i\le s$, where
		$e\ge3$ and $\ell_i\ge\lceil p/2\rceil$.
	\end{enumerate}
\end{thm}

Section~\ref{sec:prelim} collects the background used in the proofs. We prove Theorems~\ref{thm:A} and~\ref{cor:A} in Section~\ref{sec:proofA}, and Theorem~\ref{thm:B} in Section~\ref{sec:proofB}. Section 5 concludes this paper.

\section{Preliminaries}\label{sec:prelim}

Throughout this paper, let $p$ be a prime and let $q=p^f$ with $f\ge1$. We write $\Ff_q$ for the finite field with $q$ elements. Let $\Ff_q^n$ denote the $n$-dimensional vector space over $\Ff_q$, and let $\Ff_q^{k\times n}$ denote the set of all $k\times n$ matrices over $\Ff_q$. For an integer $n>1$, let $\Zz_n$ be the ring of integers modulo $n$ and let $\Zz_n^\times$ be its group of units. Let $[n]$ denote the set $\{1,2,\cdots,n\}$.

\subsection{Linear MDS codes}

Let $\cc$ be an $[n,k,d]_q$ linear code. A \emph{generator matrix} $G\in\Ff_q^{k\times n}$ of $\cc$ is a full-row-rank matrix whose row space is the code $\cc$. The \emph{dual code} of $\cc$ is defined by
\[
\cc^{\bot}=\{\bv\in\Ff_q^n: \bv\cdot \cF^{\top}=0~\mbox{for any $\cF\in \cc$}\}.
\]
Let $\E=\Ff_{q^m}$ be an extension field of $\Ff_q$. The \emph{lifted code} $\cc_{\E}$ of $\cc$ over $\E$ is given by 
\[
\cc_{\E}=\{\mathbf{u}G: \mathbf{u}\in \E^k\}.
\]
By \cite{Ding2021}, $\cc_{\E}$ is an $[n,k,d]_{q^m}$ linear code. The \emph{subfield subcode} $\cc_{\E}|_{\Ff_q}$ of $\cc_{\E}$ consists of all codewords in $\cc_{\E}$ whose coordinates lie in $\Ff_q$. The subfield subcode satisfies $\cc=\cc_{\E}|_{\Ff_q}$.

%

The following lemma shows that lifting a code commutes with taking its dual. 

\begin{lem}\label{lifted-dual}
Let $\E=\Ff_{q^m}$ be an extension field of $\Ff_q$. Let $\cc$ be an $[n,k,d]_q$ linear code and let $\cc_{\E}$ be the lifted code of $\cc$ over $\E$. Then we have $(\cc^{\bot})_{\E}=(\cc_{\E})^{\bot}$.
\end{lem}
\begin{proof}
Let $H$ be a parity-check matrix of $\cc$. Then $H$ is a generator matrix of $\cc^{\bot}$. The code $(\cc^{\bot})_{\E}$ is defined as 
\[
(\cc^{\bot})_{\E}=\{\mathbf{u}H: \mathbf{u}\in \E^{n-k}\}.
\]
It is clear that $\mathbf{u}H \cdot (\mathbf{v}G)^{\top}=\mathbf{0}$ as $HG^{\top}$ is the zero matrix. Then we have $(\cc^{\bot})_{\E}\subseteq (\cc_{\E})^{\bot}$. Both spaces have dimension $n-k$ over $\E$, so $(\cc^{\bot})_{\E}=(\cc_{\E})^{\bot}$.
\end{proof}

We recall two basic properties of MDS codes.

\begin{lem}{\rm \cite[Chapter~11]{MS}} \label{lem:mds}
Let $\cc$ be an $[n,k]_q$ MDS code. Then the following two statements hold.
\begin{enumerate}
\item Every $k$ columns of a generator matrix of $\cc$ are linearly independent over $\Ff_q$.
\item The dual of $\cc$ is MDS.
\end{enumerate}
\end{lem}

We also recall three length bounds for linear MDS codes.

\begin{lem}{\rm \cite{DFV,MS}}\label{MDS-2q-2}
A nontrivial $[n,k]_q$ MDS code satisfies
\begin{equation}\label{eq:elementarylength}
n\le q+\min\{k,n-k\}-1\le2q-2.
\end{equation}
\end{lem}

\begin{lem}{\rm \cite{Ball}}\label{lem:knownfields}
Every nontrivial linear MDS code over a prime field $\Ff_p$ has length at most $p+1$.
\end{lem}

\begin{lem}{\rm \cite{BallDeBeule}}\label{lem:BD}
	Let $p$ be an odd prime and let $q=p^f$ with $f\ge2$. Let $\cc$ be an $[n,k]_q$ MDS code. If $2\le k\le2p-2$, then we have $n\leq q+1$.
\end{lem}

To derive a further bound from Lemma~\ref{lem:BD}, we use shortening. Let $S$ be a subset of $[n]$. The \emph{punctured code} $\cc(S)$ is obtained by deleting the coordinates indexed by $S$ from each codeword of $\cc$. Let $X$ be the set of codewords of $\cc$ with entries $0$ in the positions indexed by $S$. The \emph{shortened code} $\cc^S$ is obtained by puncturing $X$ on $S$.

\begin{lem}\label{lem:shortBD}
Let $p$ be an odd prime and let $q=p^f$ with $f\ge2$. Let $\cc$ be a nontrivial $[n,k]_q$ MDS code. If $n>q+1$, then we have $n\le2q-4p+6$.
\end{lem}
\begin{proof}
Since $\cc$ is a nontrivial MDS code, it follows from Lemma \ref{lem:mds} that $\cc^{\bot}$ is a nontrivial $[n,n-k]_q$ MDS code. Then there exists a nontrivial $[n,\kappa]_q$ MDS code $\cD$ with $\kappa\leq n/2$. In fact, $\cD=\cc$ or $\cD=\cc^{\bot}$. According to Lemma \ref{lem:BD} and $n>q+1$, we have $\kappa>(2p-2)$. 

Let $S$ be a subset of $[n]$ with size $\kappa-(2p-2)$. Since $\kappa-(2p-2)<n-\kappa+1$, it follows from \cite[Theorem 1.5.7]{HP} that the shortened code $\cD^S$ has parameters $[n-\kappa+(2p-2),2p-2,n-\kappa+1]_q$. By Lemma \ref{lem:BD}, we have $n-\kappa+(2p-2)\leq q+1$. Thanks to $\kappa\leq n/2$, we have $n\le2q-4p+6$.
\end{proof}

\subsection{Cyclic codes}

Let $\gcd(n,q)=1$. An $[n,k,d]_q$ linear code $\cc$ is \emph{cyclic} if $(c_0,c_1,\cdots,c_{n-1})\in \cc$ implies that $(c_{n-1},c_0,c_1,\cdots,c_{n-2})\in \cc$. We identify the codeword $(c_0,\ldots,c_{n-1})$ with the polynomial
\[
c_0+c_1x+c_2x^2+\cdots+c_{n-1}x^{n-1}\in \Ff_q[x]/\langle x^n-1\rangle.
\]
Under this identification, the cyclic code $\cc$ is an ideal of the quotient ring $\Ff_q[x]/\langle x^n-1\rangle$. Since every ideal of $\Ff_q[x]/\langle x^n-1\rangle$ is principal, the code $\cc$ is generated by a monic polynomial $g(x)$. When $g(x)$ has the smallest degree among all generators of $\cc$, it is the \emph{generator polynomial} of $\cc$ and we write $\cc=\langle g(x)\rangle$. The dimension of $\cc$ is $n-\deg(g(x))$.

Let $m$ be the smallest positive integer such that $q^{m}\equiv 1 \pmod n$. We denote $m$ by ${\rm ord}_n(q)$. Let $\alpha$ be a primitive element of $\E=\Ff_{q^m}$ and let $\beta=\alpha^{(q^{m}-1)/n}$.
Then $\beta$ is a primitive $n$-th root of unity in $\E$.

Let $\cc$ be a cyclic code of length $n$ with generator polynomial $g(x)$. The \textit{defining set} of $\cc$ is
$
T=\{0\le i\le n-1 : g(\beta^{i})=0\}\subseteq \Zz_n.
$
By \cite[Theorem 4.4.9]{HP}, the defining set of $\cc^{\bot}$ is $\Zz_n\setminus(-T)$, where $-T=\{-i: i\in T\}$. This leads to the following definition.

\begin{defn}
	Let $\cc$ be an $[n,k,d]_q$ cyclic code with defining set $T$. The set $D=\Zz_n\setminus(-T)$ is said to be the \emph{Fourier support} of $\cc$. 
\end{defn}

Each cyclic code has a unique Fourier support $D$ such that $|D|=k$ and $qD=D$, where $qD=\{qi: i\in D\}$. The dual code $\cc^{\bot}$ has the Fourier support $\Zz_n\setminus(-D)=T$. Given $d\in\Zz_n$, we let 
\begin{equation}\label{vector-van}
\bv_d=\left(1,\beta^{d},\cdots,\beta^{(n-1)d}\right)
\end{equation}
 be a vector of length $n$ over $\E$. The set $\{\bv_d: d\in \Zz_n\}$ forms an $\E$-basis of $\E^n$. Define
\[
\cB={\rm Span}_{\E}\{\bv_d:d\in D\}.
\]
This is the linear code spanned by $\{\bv_d:d\in D\}$ over $\E$.
The following lemma identifies $\cB$ with the lifted code $\cc_{\E}$ of $\cc$ over $\E$.
\begin{lem}\label{cyclic-lifted}
Let $\cc$ be an $[n,k,d]_q$ cyclic code with Fourier support $D$. Let $\cB={\rm Span}_{\E}\{\bv_d:d\in D\}$. Then we have $\cc_{\E}=\cB$.
\end{lem}
\begin{proof}
Since $\E$ is an extension field of $\Ff_q$ and $\{\bv_d: d\in \Zz_n\}$ is an $\E$-basis of $\E^n$, each codeword $\cF$ of $\cc$ has the representation
\begin{equation}\label{lem-11-eq-1}
\cF=\sum_{x\in \Zz_n}a_x(\cF)\bv_x,~~\mbox{with}~~a_x(\cF)\in\E.
\end{equation}
Using the Fourier transform, we have $a_x(\cF)=\frac{1}{n}\bv_{-x}\cdot \cF^{\top}$. Let $T$ be the defining set of $\cc$. It is known that $\bv_d\cdot \cF^{\top}=0$ for each $\cF\in \cc$ and $d\in T$. Then we have $a_x(\cF)=0$ if $x\in -T$. By \eqref{lem-11-eq-1}, we have $\cF=\sum_{x\in D}a_x(\cF)\bv_x$. This implies that $\cc\subseteq \cB$. According to the definition of lifted codes, we have $\cc_{\E}\subseteq \cB$. Since $\cc$ is an $[n,k,d]_q$ cyclic code, its lifted code $\cc_{\E}$ has dimension $k$ over $\E$. It is clear that the code $\cB$ has dimension $k$ over $\E$. Therefore, we have $\cc_{\E}=\cB$. 
\end{proof}

We use the following notion of balance to study the Fourier support of a cyclic MDS code.

\begin{defn}
Let $t$ be a positive integer with $t\mid n$. A set $D\subseteq\Zz_n$ of size $k$ is \emph{balanced modulo $t$} if each residue class modulo $t$ contains either $\left\lfloor k/t\right\rfloor$ or $\left\lceil k/t\right\rceil$ elements of $D$. The set $D$ is said to be \emph{balanced} if this holds for every divisor $t$ of $n$.
\end{defn}

Lemma~\ref{lem:balance} shows that the Fourier support of a nontrivial cyclic MDS code is balanced modulo every divisor of its length. This is a finite-field analogue of the necessary condition established over the complex numbers in \cite[Theorem~9]{ACM}.

\begin{lem}\label{lem:balance}
The Fourier support of a nontrivial cyclic MDS code over $\Ff_q$ is balanced. 
\end{lem}
\begin{proof}
Let $t\mid n$ and $n=t\ell$. Let $D$ be the Fourier support of a nontrivial $[n,k]_q$ cyclic MDS code $\cc$. For a residue $j$ modulo $t$, write
\[
D_j=\{d\in D:d\equiv j\pmod t\}=\{j+tb:b\in B_j\},
~~ B_j\subseteq\Zz_\ell.
\]
Let $a_j=|D_j|$ for each $j\in\Zz_t$. Let $\E=\Ff_{q^m}$ be the smallest extension field of $\Ff_q$ containing a
primitive $n$th root of unity. Let $\cc_{\E}$ be the lifted code of $\cc$. 

We next prove that $a_j\le \left\lceil k/t\right\rceil$. It is clear that $0\le \left\lceil k/t\right\rceil$. We assume that $a_j>0$ for some $j\in\Zz_t$. Then the subcode
\[
W_j={\rm Span}_{\E}\{\bv_{j+tb}:b\in B_j\}\subseteq \cc_{\E}
\]
has parameters $\left[n,a_j,d_{\min}(W_j)\right]_{q^m}$, where $\bv_{j+tb}$ is the vector defined by \eqref{vector-van}. For each $\mathbf u=(u_0,\cdots,u_{n-1})\in W_j$, since $\beta$ is a primitive $n$th root of unity, we have 
\[
u_{x+y\ell}=\beta^{yj\ell}u_x,\qquad 0\le x<\ell,\quad 0\le y<t.
\]
Hence, the vectors $(u_x)_{x\in \Zz_\ell},(u_{x+\ell})_{x\in \Zz_\ell},\cdots,(u_{x+(t-1)\ell})_{x\in \Zz_\ell}$ have the same Hamming weight. Let $S=\{\ell,\ell+1,\cdots,n-1\}$. Then the punctured code $W_j(S)$ has parameters $\left[\ell,a_j,d_{\min}(W_j(S))\right]_{q^m}$ with $d_{\min}(W_j)=td_{\min}(W_j(S))$. According to the Singleton bound on $W_j(S)$, we have 
\begin{equation}\label{lem-12-eq-1}
d_{\min}(W_j)=td_{\min}(W_j(S))\leq t(\ell-a_j+1).
\end{equation}
Since $\cc_{\E}$ is an $[n,k]_{q^m}$ MDS code and $W_j\subseteq \cc_{\E}$, we have
\begin{equation}\label{lem-12-eq-2}
	d_{\min}(W_j)\ge n-k+1.
\end{equation}
Combining \eqref{lem-12-eq-1} and \eqref{lem-12-eq-2}, we have $a_j\le \left\lceil k/t\right\rceil$. 

For the lower bound, we prove $a_j\ge \left\lfloor k/t\right\rfloor$. The Fourier support of $\cc^{\bot}$ is $\Zz_n\setminus(-D)$. Let 
\[
T_j=\{d\in \Zz_n\setminus(-D):d\equiv -j\pmod t\}.
\]
Since $a_j=|D_j|$, we have $|T_j|=\ell-a_j$. Applying the same argument to the dual code, we have $\ell-a_j\le \left\lceil (n-k)/t\right\rceil$. This implies that $a_j\ge \left\lfloor k/t\right\rfloor$. Therefore, the Fourier support $D$ is balanced. 
\end{proof}

The following lemma gives necessary conditions for a nontrivial $[n,k]_q$ cyclic MDS code to have length $n>q+1$. We use these conditions in the proof of Theorem~\ref{thm:B}.

\begin{lem}\label{lem:descent}
Let $p$ be a prime and let $f$ be a positive integer. Let $q=p^f$ and $t={\rm ord}_n(p)$. Let $\cc$ be a nontrivial $[n,k]_q$ cyclic MDS code with $n>q+1$. Then we have $f\mid t$ and ${\rm ord}_n(q)=t/f$.
\end{lem}
\begin{proof}
	Put $h=\gcd(f,t)$ and let $g(x)$ be the monic generator
	polynomial of $\mathcal C$ with degree $n-k$. Since $p^t\equiv1\pmod n$, all
	$n$-th roots of unity belong to $\mathbb F_{p^t}$. As
	$g(x)\mid x^n-1$, its coefficients lie in $\mathbb F_{p^t}$.
	These coefficients also lie in $\mathbb F_{p^f}$. Therefore,
	\[
	g(x)\in
	\bigl(\mathbb F_{p^f}\cap\mathbb F_{p^t}\bigr)[x]
	=\mathbb F_{p^h}[x].
	\]
	
	Let $\mathcal C_0$ be the cyclic code of length $n$ over
	$\mathbb F_{p^h}$ generated by $g(x)$. Furthermore, we have $\mathcal C_0\subseteq\mathcal C$, which implies that every nonzero codeword of $\mathcal C_0$ has weight at
	least $n-k+1$. By the Singleton bound, $\mathcal C_0$ is
	a nontrivial $[n,k]_{p^h}$ MDS code. By Lemma \ref{MDS-2q-2}, we have $n\leq 2p^h-2$.
	
	Assume by contradiction that $f\nmid t$. Then we obtain $h<f$ and $2h\le f$. It follows that
	\[
	n\le2p^h-2<p^{2h}\le p^f=q,
	\]
	contradicting $n>q+1$. Therefore, we have $f\mid t$. It is easy to check that 
	\[
	\operatorname{ord}_n(q)
	=\operatorname{ord}_n(p^f)
	=\frac{\operatorname{ord}_n(p)}
	{\gcd(\operatorname{ord}_n(p),f)}
	=\frac{t}{f}.
	\]
\end{proof}

\subsection{Locally recoverable codes}

We use locally recoverable codes in the proof of Theorem~\ref{thm:A}. The following definition fixes the notation.

\begin{defn}\label{Definition-LRC}{\rm \cite{GHYS}}
	An $[n,k,d]_q$ linear code $\cc$ has \emph{locality} $r$ if for each $i\in[n]$ and each codeword $\mathbf{c}=(c_1,\cdots,c_{n})\in\cc$ there exists a linear function $\rho_i:\Ff_q^r\rightarrow \Ff_q$ such that $c_i=\rho_i(c_{t_1},\cdots,c_{t_r})$ with $t_1,\cdots,t_r\in[n]\setminus\{i\}$.
\end{defn}

By \cite{GHYS}, every $[n,k,d]_q$ linear code with locality $r$ satisfies the Singleton-like bound
\begin{equation}\label{C-Singleton}
	d\leq n-k-\left\lceil\frac{k}{r}\right\rceil+2.
\end{equation}

\section{Proof of Theorems \ref{thm:A} and \ref{cor:A}}\label{sec:proofA}

We prove Theorems~\ref{thm:A} and~\ref{cor:A} using the auxiliary lemmas below.

\subsection{Auxiliary lemmas}

The key step is to bound the number of residues fixed by a nonidentity multiplier that preserves the Fourier support. We obtain this bound from the locality of an auxiliary code.

\begin{lem}\label{lem:fixed}
Let $D$ be the Fourier support of a nontrivial $[n,k]_q$ cyclic MDS code. Let $\Zz_n^\times$ be the unit group of $\Zz_n$. For every $u\in\Zz_n^\times$ satisfying $uD=D$ and $u\ne1$, one has $\gcd(u-1,n)\le2$.
\end{lem}
\begin{proof}
We first prove the case that $u$ has prime order $r$ in $\Zz_n^\times$. Given $j\in \Zz_n$, let 
\[
O_j=\{j,ju,\cdots,ju^{\ell_j-1}\}\mod n
\] 
be an orbit of multiplication by $u$ on $\Zz_n$. Since $u$ has prime order $r$, we have $\ell_j=1$ or $\ell_j=r$ for each $j\in \Zz_n$. If $\ell_j=1$, then we have $(u-1)j\equiv0\pmod n$ which implies that $\frac{n}{\gcd(u-1,n)}\mid j$. Then there are $f=\gcd(u-1,n)$ orbits with size $1$. Let $b$ be the number of orbits with size $r$. Then we have $n=f+br$. Due to $uD=D$, the set $D$ is the disjoint union of some $O_j$. Assume that $D$ has $a$ orbits of size $1$ and $t$ orbits of size $r$. Then we have $k=a+tr$. 

Let $\cc$ be a nontrivial $[n,k]_q$ cyclic MDS code. Let $m={\rm ord}_n(q)$	and let $\E=\Ff_{q^m}$. Let $\bv_d$ be the vector over $\E$ defined by \eqref{vector-van}. By Lemma \ref{cyclic-lifted}, the lifted code $\cc_{\E}$ is an $[n,k]_{q^m}$ MDS code spanned by $\{\bv_d:d\in D\}$ over $\E$. Let $D=\{d_1,\cdots,d_k\}$. The code $\cc_{\E}$ has a generator matrix
\[
G=\begin{pmatrix}
	1 & \beta^{d_1} &  \cdots     &  \beta^{(n-1)d_1} \\
	\vdots	& \vdots	& \ddots  &  \vdots \\
	1 & \beta^{d_k} &  \cdots     &  \beta^{(n-1)d_k}
\end{pmatrix}.
\]

If $t=0$, then each element in $D$ is a multiple of $n/f$. Hence $\beta^{df}=1$ for each $d\in D$. This contradicts the fact that any $k$ columns of $G$ are linearly independent over $\E$ as $\cc_{\E}$ is MDS. Then we have $t>0$. By Lemma \ref{lem:mds}, the dual code $\cc_{\E}^{\bot}$ is MDS and has the Fourier support $\Zz_n\setminus(-D)$. It has $f-a$ orbits of size $1$ and $b-t$ orbits of size $r$. Using the same idea as shown before, we have $t<b$. Therefore, we have $0<t<b$.

We define a linear map $P:\E^n\longrightarrow \E^n$ by $(P\mathbf x)_j=x_{uj}$ with ${\mathbf x}=(x_0,\cdots,x_{n-1})\in \E^n$. It is clear that $P\bv_d=\bv_{ud}$. Let $P\cc_{\E}=\{P\cF:\cF\in\cc\}$. Then we have 
$
\cc_{\E}=P\cc_{\E}.
$
Let $I$ denote the identity linear map on $\E^n$. We construct a linear subcode $\cD=(P-I)\cc_{\E}$ of $\cc_{\E}$. Given $d\in D$, if $d\equiv ud \pmod n$, then we derive that $(P-I)\bv_d=\mathbf 0$. On the orbit $O_d$ of size $r$ in $D$, let $\mathbf a=\sum_{i=0}^{r-1}y_i \bv_{u^id}$ with $y_i\in \E$. Then we have
\begin{equation}\label{lem-18-eq-1}
(P-I)\mathbf a=\sum_{i=0}^{r-1}(y_i-y_{i+1})\bv_{u^{i+1}d}, 
\end{equation}
where the indices of $y_i$ are taken modulo $r$. Let $\cB_d={\rm Span}_{\E}\{\bv_d:d\in O_d\}$. Then the code $(P-I)\cB_d$ has dimension $r-1$ as $\bv_{d},\bv_{ud},\cdots,\bv_{u^{r-1}d}$ are linearly independent over $\E$. Note that $D$ is the disjoint union of $a$ orbits of size $1$ and $t$ orbits of size $r$. Since $(P-I)\bv_d=\mathbf 0$ if $d\equiv ud \pmod n$, the dimension of $\cD$ is $(r-1)t$.

We next prove that the code $\cD$ has locality $r-1$. Let $\mathbf x=(P-I)\cF$ with $\cF=(c_0,\cdots,c_{n-1})\in \cc$ be a codeword of $\cD$. Then the $j$-th component $x_j$ of $\mathbf x$ is $c_{uj}-c_j$. Given $d\in D$ with $d\equiv ud \pmod n$, we have $x_d=c_{ud}-c_d=0$. Since there are $f$ orbits with size $1$ in $\Zz_n$, the components indexed by these orbits of $\mathbf x$ are zero. On an orbit $O$ with size $r$ of $\Zz_n$, we have
\begin{equation}\label{eq:zero-sum}
	\sum_{j\in O}x_j
	=\sum_{j\in O}c_{uj}-\sum_{j\in O}c_j=0.
\end{equation}
Let $S=\{d\in \Zz_n: d\equiv ud \pmod n\}$. The punctured code $\cD(S)$ has parameters $[n-f=br,(r-1)t,d(\cD)]_{q^m}$ with $d(\cD)$ being the minimum distance of $\cD$. According to Definition \ref{Definition-LRC} and \eqref{eq:zero-sum}, the code $\cD(S)$ has locality $r-1$. By the bound in \eqref{C-Singleton}, we have
\begin{equation}\label{eq-Udistance}
d(\cD)\le r(b-t)+2.
\end{equation}
Since $\cc_{\E}$ is an $[n,k]_{q^m}$ MDS code and $\cD$ is a subcode of $\cc_{\E}$, we have 
\begin{equation}\label{eq:Udistance-1}
	d(\cD)\ge n-k+1=f-a+(b-t)r+1.
\end{equation}
Combining \eqref{eq-Udistance} and \eqref{eq:Udistance-1}, we have $f-a\le 1$.

Now replace $\cc_{\E}$ by $(\cc^\bot)_\E$. By Lemma \ref{lifted-dual}, we have $(\cc^\bot)_\E=\cc_\E^\bot$. It has the Fourier support $\Zz_n\setminus(-D)$. Since $D$ has $a$ orbits of size $1$ and $t$ orbits of size $r$, $\Zz_n\setminus(-D)$ is the disjoint union of $f-a$ orbits of size $1$ and $b-t$ orbits of size $r$. Using the similar technique as shown above, we derive that $f-(f-a)\le 1$ which implies that $a\leq 1$. By $f-a\le 1$, we have $f=\gcd(u-1,n)\le 2$.
	
We prove the case that $u\in \Zz_n^\times$ have arbitrary order $h>1$. Let $r$ be a prime with $r\mid h$ and let $v=u^{h/r}$. Since $v$ has prime order $r$ and $vD=D$, the prime-order case gives that $\gcd(v-1,n)\leq 2$. Since $(u-1)j\equiv0\pmod n$ implies that $(v-1)j\equiv0\pmod n$, every orbit of size one under multiplication by $u$ is also an
orbit of size one under multiplication by $v$. Since these two actions have
$\gcd(u-1,n)$ and $\gcd(v-1,n)$ orbits of size one, respectively,
we obtain 
\[
\gcd(u-1,n)\leq\gcd(v-1,n)\leq 2.
\]
This completes the proof.	
\end{proof}

Using Lemma~\ref{lem:fixed}, we prove the MDS length bound for nontrivial $q$-ary cyclic MDS codes with $n\mid (q^2-1)$.

\begin{lem}\label{lem:quadratic}
If a nontrivial $[n,k]_q$ cyclic MDS code satisfies $n\mid (q^2-1)$, then $n\le q+1$. 
\end{lem}
\begin{proof}
If $q\equiv1\pmod n$, then $n\mid q-1$ and the conclusion is immediate. Otherwise $q$ has order $2$ modulo $n$. Thanks to $qD=D$, it follows from Lemma~\ref{lem:fixed} that $f=\gcd(q-1,n)\le2$.

Let $n=fn_0$ and $q-1=fb$. By $n\mid (q^2-1)$, we have $n_0\mid (q+1)$. Since $f\leq 2$, we have $n\mid 2(q+1)$. By Lemma \ref{MDS-2q-2}, we have $n\leq 2q-2<2q+2$. Then we have $n\le q+1$ as $n\mid 2(q+1)$.
\end{proof}

Applying the multiplier bound in Lemma~\ref{lem:fixed} to reduction modulo a divisor gives the following equality of multiplicative orders.

\begin{lem}\label{cor:divisor}
Let $\cc$ be a nontrivial $[n,k]_q$ cyclic MDS code. Let $m\mid n$ with $m\ge3$. Then we have ${\rm ord}_n(q)={\rm ord}_m(q)$. 
\end{lem}
\begin{proof}
Let $D$ be the Fourier support of $\cc$ and let $H(D)=\{u\in\Zz_n^\times: uD=D\}$. We define the map $\tau$ by
\[
H(D)\longrightarrow\Zz_m^\times,\quad u\longmapsto u\pmod m.
\]
Since $H(D)$ is a subgroup of $\Zz_n^\times$ and $\tau$ is the
restriction of the natural reduction map, $\tau$ is a group
homomorphism. It remains to prove that $\tau$ is injective. Let $u\in\ker\tau$. Then $uD=D$ and $m\mid u-1$.
Since $m\mid n$, we have $m\mid\gcd(u-1,n)$.
If $u\not\equiv1\pmod n$, Lemma~\ref{lem:fixed} implies
\[
3\le m\le\gcd(u-1,n)\le2,
\]
a contradiction. Hence $u\equiv1\pmod n$ which says that $\ker\tau=\{1\}$. Therefore $\tau$ is injective.

Let $h={\rm ord}_m(q)$ and $t={\rm ord}_n(q)$. Due to $m\mid n$, we have $h\mid t$. By the definition of $\cc$, we have $q\in H(D)$. Then we have $q^h\in H(D)$. By $q^h\equiv1\pmod m$ and $\ker\tau=\{1\}$, we obtain $q^h\equiv1\pmod n$. This implies that $t\mid h$. Therefore, we have ${\rm ord}_n(q)={\rm ord}_m(q)$.
\end{proof}

Lemmas~\ref{lem:fixed}, \ref{lem:quadratic}, and \ref{cor:divisor}
yield Theorem~\ref{thm:A} under condition~(1).
The following lemma handles condition~(2) and is also used in the proof of Theorem~\ref{thm:B}. 

\begin{lem}\label{lem:boundary}
Let $\cc$ be a nontrivial $[n,k]_q$ cyclic MDS code such that $\gcd(n,q^2-1)=2$ and let $m=\ord_n(q)$. Then we have
\[
k\equiv n-k\equiv1\pmod{2m}~~\hbox{and}~~\min\{k,n-k\}\ge2m+1.
\]
\end{lem}
\begin{proof}
Since $\gcd(n,q^2-1)=2$ and $\gcd(n,q)=1$, we obtain $n$ is even and $q$ is odd. Thus $4\mid q^2-1$ and the assumption $\gcd(n,q^2-1)=2$ shows that $n/2$ is odd and $m>2$.
	
	Let $D$ be the Fourier support of $\cc$.
	Choose any prime divisor $r$ of $m$ and let
	$u=q^{m/r}\pmod n$.
	Then $u$ has order $r$ in $\Zz_n^\times$ and satisfies $uD=D$.
	By Lemma~\ref{lem:fixed}, we have $\gcd(u-1,n)\le2$.
	Since $u$ is odd and $n$ is even, it follows that
	$\gcd(u-1,n)=2$.
	
	Given $j\in\Zz_n$, let
	\[
	O_j=\{j,ju,\ldots,ju^{\ell_j-1}\}\pmod n
	\]
	be its orbit under multiplication by $u$ in $\Zz_n$.
	Since $u$ has prime order $r$, we have $\ell_j\in\{1,r\}$.
	The equality $\gcd(u-1,n)=2$ shows that the only orbits
	of size one are $O_0=\{0\}$ and $O_{n/2}=\{n/2\}$.
	Let $a=|D\cap\{0,n/2\}|$.
	The prime-order argument in the proof of
	Lemma~\ref{lem:fixed} gives $a\le1$ and $\gcd(u-1,n)-a\le1$. Hence, we have $a=1$.
	
	Next consider the orbits $\Omega_j$ under multiplication by $q$ in $\Zz_n$.
	For each $j\in\Zz_n\setminus\{0,n/2\}$, we have
	$n/\gcd(n,j)\ge3$.
	The orbit $\Omega_j$ has length
	\[
	\ord_{n/\gcd(n,j)}(q)=\ord_n(q)=m,
	\]
	where the equality follows from Lemma~\ref{cor:divisor}.
	Thus, apart from $\{0\}$ and $\{n/2\}$, all orbits $\Omega_j$
	have size $m$.
	Since $q$ is odd, each such orbit consists entirely
	of even integers or entirely of odd integers.
	
	Let $x$ and $y$ denote the numbers of these two types
	of orbits $\Omega_j$ contained in $D$, respectively.
	Let $\epsilon=1$ if $0\in D$ and $\epsilon=0$ otherwise.
	Since $n/2$ is odd and $a=1$, the numbers of even and odd
	elements of $D$ are $mx+\epsilon$ and $my+1-\epsilon$, respectively.
	By Lemma~\ref{lem:balance}, these numbers differ by at most one.
	Hence
	\[
	|m(x-y)+2\epsilon-1|\le1.
	\]
	Since $m>2$, this forces $x=y$ and therefore
	\[
	k=|D|=1+2mx.
	\]
	Since $\cc$ is a nontrivial MDS code, we have $k\ge2$ which implies that $x\ge1$. Thus $k\equiv1\pmod{2m}$ and $k\ge2m+1$.
	
	Finally, $\cc^\perp$ is a nontrivial $[n,n-k]_q$
	cyclic MDS code.
	Applying the same argument to $\cc^\perp$ gives
	$n-k\equiv1\pmod{2m}$ and $n-k\ge2m+1$.
	This completes the proof.
\end{proof}

\subsection{Proof of Theorem~\ref{thm:A}} 

Let $f=\gcd(n,q^2-1)$. By Lemma \ref{lem:boundary}, we have $f\neq 2$ if $k$ is even. By our assumption $f>1$ in Condition (2), we have $f>2$. Therefore $f$ is always greater than $2$ in Conditions (1) and (2) provided in Theorem \ref{thm:A}.

Let $D$ be the Fourier support of $\cc$. According to $qD=D$, we have $q^2D=D$. If $q^2\not\equiv1\pmod n$, Lemma~\ref{lem:fixed} gives $f\le2$. This contradicts our assumption $f>2$. Therefore we have $n\mid q^2-1$. By Lemma \ref{lem:quadratic}, we obtain that $n\leq q+1$. This completes the proof of Theorem \ref{thm:A}.

\subsection{Proof of Theorem~\ref{cor:A}} 

For the case (1), since $3\mid n$ or $4\mid n$, it follows from Lemma \ref{cor:divisor} that ${\rm ord}_n(q)={\rm ord}_3(q)$ or ${\rm ord}_n(q)={\rm ord}_4(q)$. Since the unit groups of $\Zz_3^{\times}$ and  $\Zz_4^{\times}$ have order $2$. Then we have ${\rm ord}_n(q)\mid 2$. By Lemma \ref{lem:quadratic}, we have $n\leq q+1$.

For the case (2), since $\ell_1\mid n$ and $\ell_2\mid n$, it follows from Lemma \ref{cor:divisor} that ${\rm ord}_n(q)={\rm ord}_{\ell_1}(q)={\rm ord}_{\ell_2}(q)$. Since $\ell_1$ and $\ell_2$ are odd primes, the unit groups of $\Zz_{\ell_1}^{\times}$ and  $\Zz_{\ell_2}^{\times}$ have respective orders $\ell_1-1$ and $\ell_2-1$. Then we have ${\rm ord}_n(q)\mid 2$ as $\gcd(\ell_1-1,\ell_2-1)=2$. By Lemma \ref{lem:quadratic}, we have $n\leq q+1$.

\section{Proof of Theorem~\ref{thm:B}}\label{sec:proofB}

We prove Theorem~\ref{thm:B} by considering three forms of the length. By Theorem~\ref{cor:A}, it suffices to consider the cases $n=\varepsilon5^{\gamma}$, $n=\varepsilon7^{\delta}$, and $n=\varepsilon\ell_i^{e_i}$, where $\varepsilon\in\{1,2\}$. The auxiliary lemmas below address these cases. Throughout this section, we write $q=p^f$ and $m=\ord_n(q)$.

\subsection{Auxiliary lemmas}

For a prime $r$ and a nonzero integer $a$, let $\nu_r(a)$ denote
the exponent of $r$ in $|a|$. We use the following classical
lifting-the-exponent identity, commonly attributed to
Carmichael \cite{Carmichael1909} and Lucas \cite{Lucas1878}.

\begin{lem}\label{lem:LTE}
	Let $r$ be a prime and let $x,y$ be distinct integers such that
	$r\nmid xy$ and $r\mid(x-y)$. Suppose that either $r$ is odd or
	$4\mid(x-y)$. Then, for every positive integer $h$,
	\[
	\nu_r(x^h-y^h)=\nu_r(x-y)+\nu_r(h).
	\]
\end{lem}

We collect the restrictions that any nontrivial cyclic MDS code of length $n>q+1$ must satisfy. Field descent provides the relation between the orders of $p$ and $q$.

\begin{lem}\label{lem:normal}
Let $\cc$ be a nontrivial $[n,k]_q$ cyclic MDS code with $n>q+1$. Let $q=p^f$ with $f$ being a positive integer.  Let $t=\ord_n(p)$ and $m=\ord_n(q)$. Then we have
\[
f>1,~q+1<n\le2q-2,~t=fm ~\mbox{and}~ m>2.
\]
\end{lem}
\begin{proof}
Lemma~\ref{lem:knownfields} excludes $f=1$ and Lemma~\ref{MDS-2q-2} gives $n\le2q-2$. By Lemma \ref{lem:descent}, we have $t=fm$. 

If $m\le2$, then $n\mid q^2-1$. Lemma~\ref{lem:quadratic} implies that $n\le q+1$, which contradicts our assumption $n>q+1$. Hence $m>2$.
\end{proof}

\begin{lem}\label{lem:five}
Let $\varepsilon\in\{1,2\}$ and let $e$ be a positive integer. Then every nontrivial $[n,k]_q$ cyclic MDS code of length $n=\varepsilon5^e$ satisfies $n\le q+1$.
\end{lem}
\begin{proof}
Assume by contradiction that $n>q+1$ and use the notation of Lemma~\ref{lem:normal}.
	By Lemma~\ref{cor:divisor}, we obtain $m=\ord_5(q)$ with $m\mid4$. Since $m>2$, we have $m=4$.
	If $\varepsilon=2$, then $p$ is odd and its order modulo two is one.
	Thus, we have
	\[
	t=\ord_{n}(p)\mid4\cdot5^{e-1}.
	\]
	As $t=4f$, we obtain $f\mid5^{e-1}$. By Lemma~\ref{lem:normal}, $f>1$ is a power
	of $5$.
	
We next prove that $\ord_{5^e}(q)$ is $4$. This is immediate if $\varepsilon=1$ as $m=4$. When $\varepsilon=2$, we have $\ord_{2}(q)=1$. Since $\gcd(2,5^e)=1$, we obtain $\ord_{n}(q)={\rm lcm}(\ord_{2}(q),\ord_{5^e}(q))$. This implies that $\ord_{5^e}(q)=4$ as $\ord_n(q)=4$. Therefore, we have $\ord_{5^e}(q)=4$ when $\varepsilon\in\{1,2\}$. This implies that $5^e\mid (p^{2f}+1)$.
	
Since $\ord_5(p^f)=4$, $\ord_5(p)\mid4$ and 
\[
\ord_5(p^f)=\frac{\ord_5(p)}{\gcd(f, \ord_5(p))},
\]
we derive that $\ord_5(p)=4$ which implies that $5\mid (p^2+1)$. Apply Lemma~\ref{lem:LTE} with $r=5$, $x=p^2$, $y=-1$ and $h=f$. Since $f$ is odd and $5^e\mid (p^{2f}+1)$, this gives
	\[
	e\le\nu_5(p^{2f}+1)=\nu_5(p^2+1)+\nu_5(f).
	\]
According to $5^{\nu_5(f)}=f$, we obtain
	\begin{equation}\label{eq:fivebound}
		n=\varepsilon5^e\le\varepsilon f(p^2+1).
	\end{equation}
	
	For $p\ge3$, the sequence $p^j/j$ is increasing for $j\ge2$ and
	$p^5>10(p^2+1)$. Since $f\ge5$, we have
	\[
	\frac{p^f}{f}\ge\frac{p^5}{5}>2(p^2+1).
	\]
	Together with \eqref{eq:fivebound}, this gives $n<q$, a contradiction.
	For $p=2$ the fact $\gcd(n,q)=1$ forces $\varepsilon=1$, and
	\eqref{eq:fivebound} gives $n\le5f$. The
	same monotonicity gives $2^f>5f$ for every $f\ge5$. Then we have $n<q=2^f$,
	a contradiction. This completes the proof.
\end{proof}

The order and valuation arguments also apply to lengths $\ell^e$ and $2\ell^e$, where $\ell=2r+1$ and both $r$ and $\ell$ are odd primes. In this case, a counterexample must be defined over a quadratic extension of the prime field.

\begin{lem}\label{lem:square}
	Let $r$ and $\ell=2r+1$ be odd primes and let
	$n=\varepsilon\ell^e$ with $\varepsilon\in\{1,2\}$ and $e\ge1$ being an integer. Let $q=p^f$ with $f$ being a positive integer.
	If a nontrivial $[n,k]_q$ cyclic MDS code satisfies $n>q+1$, then we obtain that $q=p^2$ with $p$ being odd prime, $\ord_n(q)=r$, $\ord_{\ell^e}(p)=\ord_\ell(p)=2r$ and $\ell^e\mid\frac{p^r+1}{p+1}$.
\end{lem}
\begin{proof}
Let $m=\ord_n(q)$ and $t=\ord_{n}(p)$. By Lemma~\ref{cor:divisor}, we have $m=\ord_\ell(q)$ with $m\mid2r$. According to Lemma \ref{lem:normal}, we get $m>2$. Since $r$ is a prime, we have 
\begin{equation}\label{eq:twoorders}
	m\in\{r,2r\}.
\end{equation}
It is easy to check that $t=\ord_{\ell^e}(p)$ and $t\mid 2r\ell^{e-1}$ by a similar argument used in the proof of Lemma \ref{lem:five}. By Lemma \ref{lem:normal}, we have $t=fm$ with $f>1$. It follows from \eqref{eq:twoorders} that $f\mid2\ell^{e-1}$. Hence, $f=2$ or $f\ge\ell=2r+1$. 

We next exclude the case $f\ge\ell=2r+1$. Due to $\ord_\ell(p^f)=\frac{\ord_\ell(p)}{\gcd(f, \ord_\ell(p))}$, the order $\ord_\ell(p)$ is a multiple of $m$. As $\ord_\ell(p) \mid 2r$, we have $\ord_\ell(p)=r$ or $\ord_\ell(p)=2r$. Let $v=\nu_\ell(p^{2r}-1)$. If $\ord_\ell(p)=r$, the entire $\ell$-part of $p^{2r}-1$ lies in $p^r-1$ by $\gcd(p^r-1,p^r+1)\le 2$. If $\ord_\ell(p)=2r$, the entire $\ell$-part lies in $p^r+1$ by $\gcd(p^r-1,p^r+1)\le 2$. In both cases, we have $\ell^v\le p^r+1$. Since $\ell^e\mid q^{2r}-1=p^{2rf}-1$, by
Lemma~\ref{lem:LTE} with the prime $\ell$, $x=p^{2r}$, $y=1$ and $h=f$, we get $e\le v+\nu_\ell(f)$. Consequently,
\begin{equation}\label{eq:growthbound}
n=\varepsilon\ell^e\le2f(p^r+1).
\end{equation}
Since $r\ge 3$ and $p\ge 2$, it is clear that $p^{-r}\le1/8$ and $\frac{p^{r+1}}{2r+1}\ge 16/7$. Hence, we get 
\begin{equation}\label{eq:growthcomparison}
p^{2r+1}>2(2r+1)(p^r+1).
\end{equation}	
If $f\ge2r+1$, monotonicity of $p^j/j$ for $j\ge2$ and \eqref{eq:growthcomparison} give
\begin{equation}\label{eq:growthbound-1}
	\frac{p^f}{f}\ge\frac{p^{2r+1}}{2r+1}>2(p^r+1).
\end{equation}
By \eqref{eq:growthbound} and \eqref{eq:growthbound-1}, we have $n<q$ which contradicts our assumption $n>q+1$. Therefore, we have $f=2$.
	
Now $t=2m\mid2r\ell^{e-1}$. Then we have $m=r$ as $r$ and $\ell$ are odd primes. If $p=2$, then $q=4$ and $n\ge\ell\ge7>2q-2$, contrary to Lemma~\ref{MDS-2q-2}. Hence $p$ is odd.
	
Since $\ord_{\ell^e}(p)=t=2m=2r$, the element $p^r$ has order
two modulo $\ell^e$. Then we obtain
\[
p^r\equiv-1\pmod{\ell^e}.
\]
In particular, $p^r\not\equiv1\pmod\ell$, since $\ell$ is odd.
As $\ord_\ell(p)\in\{r,2r\}$, established earlier,
it follows that $\ord_\ell(p)=2r$.
Thus $\ell\nmid p+1$, since otherwise
$\ord_\ell(p)\le2$.
Consequently, $\gcd(\ell^e,p+1)=1$ and
$\ell^e\mid p^r+1$ implies
$
\ell^e\mid\frac{p^r+1}{p+1}
$
This completes the proof.
\end{proof}

\begin{lem}\label{lem:seven}
Every nontrivial $[n,k]_q$ cyclic MDS code of length $n=\varepsilon7^e$, where $\varepsilon\in\{1,2\}$ and $e\ge1$ is an integer, satisfies $n\le q+1$.
\end{lem}
\begin{proof}
Suppose by contradiction that $n>q+1$. Applying Lemma~\ref{lem:square} with $r=3$ and
	$\ell=7$, we have $q=p^2$, $p$ is odd and
	\[
	7^e\mid\frac{p^3+1}{p+1}=p^2-p+1.
	\]
	If $\varepsilon=1$, this gives $n\le p^2-p+1<q+1$, a contradiction.
	Thus $\varepsilon=2$ and $n=2\cdot7^e$.
	
	Let $j=(p^2-p+1)/7^e$. By our assumption $n=2\cdot7^e>p^2+1$, we have
	\[
	1\le j<\frac{2(p^2-p+1)}{p^2+1}<2.
	\]
	Hence $j=1$ and $n=2(p^2-p+1)$.
	By Lemma \ref{lem:shortBD}, we get $n\le2p^2-4p+6$, whereas
	\[
	2(p^2-p+1)-(2p^2-4p+6)=2p-4>0.
	\]
	This is a contradiction. Therefore, we have $n\le q+1$. 
\end{proof}

The cases of length $2\ell$ and $2\ell^2$ also require information about the dimension. The following lemma gives a divisibility condition for cyclic MDS codes whose length is twice their dimension. 

\begin{lem}\label{lem:half}
Let $\cc$ be a $[2h,h]_q$ cyclic MDS code with $h\ge2$. Then we have $2h\mid q^2-1$.
\end{lem}
\begin{proof}
Let $m={\rm ord}_n(q)$	and let $\E=\Ff_{q^m}$. Let $\beta$ be a primitive $n$-th root of unity in $\E$ and let $\bv_d$ be the vector over $\E$ defined by \eqref{vector-van}. Let $D$ be the Fourier support of $\cc$. By Lemma \ref{cyclic-lifted}, the lifted code $\cc_{\E}$ is a $[2h,h]_{q^m}$ MDS code spanned by $\{\bv_d:d\in D\}$ over $\E$. It is clear that $\cc_{\E}$ is an MDS cyclic code. Let
\[
\cc^{\prime}_{\E}=\{(c_0,c_2,\cdots,c_{2h-2},c_1,c_3,\cdots,c_{2h-1}):(c_0,\cdots,c_{2h-1})\in \cc_{\E}\}.
\]
The code $\cc^{\prime}_{\E}$ is a $[2h,h]_{q^m}$ MDS code. Since $\cc_{\E}$ is cyclic, for each 
$$
(c_0,c_2,\cdots,c_{2h-2},c_1,c_3,\cdots,c_{2h-1})\in \cc^{\prime}_{\E},
$$
we have $(c_{2h-2},c_0,\cdots,c_{2h-4},c_{2h-1},c_1,\cdots,c_{2h-3})\in \cc^{\prime}_{\E}$. Then the code $\cc^{\prime}_{\E}$ has a generator matrix
\begin{equation}\label{lem27-eq-2}
G=(I_h\mid A)~~\mbox{with}~~	
A=\begin{pmatrix}
	a_0     & a_1     & \cdots & a_{h-1}\\
	a_{h-1} & a_0     & \cdots & a_{h-2}\\
	\vdots  & \vdots  & \ddots & \vdots\\
	a_1     & a_2     & \cdots & a_0
\end{pmatrix},
\end{equation}
for some $a_0,\cdots,a_{h-1}\in \E$. We also note that every $2\times2$ minor of $A$ is nonzero.
	Indeed, take any two rows and two columns of $A$.
	Adjoin the $h-2$ columns of $I_h$ indexed by the remaining rows.
	The resulting $h\times h$ submatrix of $G$ has determinant equal,
	up to sign, to the chosen $2\times2$ minor.
	Its determinant is nonzero as any $h$ columns of $G$ are linearly independent over $\E$.

Since $\gcd(2h,q)=1$, $q$ is odd and let $q=2s+1$. For $d\in\Zz_{2h}$, define
\[
\bv'_d=(\bv'_{d,1},\bv'_{d,2}),
~~\mbox{with}~~
\bv'_{d,1}=(\beta^{2jd})_{j\in\Zz_h},
~~
\bv'_{d,2}=(\beta^{(2j+1)d})_{j\in\Zz_h}.
\]
By the definition of $\cc'_{\E}$, this code is spanned by $\{\bv'_d:d\in D\}$ over $\E$. Define two $\E$-linear maps on $\E^h$ by
\[
(P_q\mathbf c)_j=c_{qj},
~~
(Q_q\mathbf c)_j=c_{qj+s},
\]
where $\cF=(c_0,\cdots,c_{h-1})$ and the indices are taken modulo $h$.
Using $q=2s+1$, we obtain
\[
(P_q\bv'_{d,1})_j
=\beta^{2qjd}
=(\bv'_{qd,1})_j
\]
and
\[
(Q_q\bv'_{d,2})_j
=\beta^{(2(qj+s)+1)d}
=\beta^{q(2j+1)d}
=(\bv'_{qd,2})_j.
\]
Hence
\[
(P_q\bv'_{d,1},Q_q\bv'_{d,2})=\bv'_{qd}.
\]
Since $qD=D$, it follows that
\begin{equation}\label{lem27-eq-1}
	\cc'_{\E}
	=
	\left\{
	(P_q\mathbf c_1,Q_q\mathbf c_2):
	(\mathbf c_1,\mathbf c_2)\in\cc'_{\E}
	\right\}.
\end{equation}

By \eqref{lem27-eq-2}, we have $(\mathbf e_0,(a_j)_{j\in\Zz_h})\in 	\cc'_{\E}$ with $\mathbf e_0=(1,0,\cdots,0)$. It follows from \eqref{lem27-eq-1} that
\[
(P_q\mathbf e_0,Q_q(a_j)_{j\in\Zz_h})=(\mathbf e_0,(a_{qj+s})_{j\in\Zz_h})\in\cc'_{\E},
\]
where the indices are taken modulo $h$. Since the minimum distance of $\cc'_{\E}$ is $h+1$, we have $a_{qj+s}=a_j$ for each $j\in\Zz_h$. Taking $j=0$ and $j=h-1$, respectively, yields
\[
a_s=a_0~~\mbox{and}~~a_{q(h-1)+s}=a_{h-1}.
\]
Since $q=2s+1$, we have $a_{q(h-1)+s}=a_{h-s-1}=a_{h-1}$. Suppose that neither $h\mid s$ nor $h\mid s+1$.
Then rows $0$ and $s+1\pmod h$ of $A$ are distinct, as are
columns $0,s\pmod h$.
The corresponding submatrix of $A$ is
\[
\begin{pmatrix}
	A_{0,0}   & A_{0,s}\\
	A_{s+1,0} & A_{s+1,s}
\end{pmatrix}
=
\begin{pmatrix}
	a_0      & a_s\\
	a_{h-s-1} & a_{h-1}
\end{pmatrix}
=
\begin{pmatrix}
	a_0    & a_0\\
	a_{-1} & a_{-1}
\end{pmatrix}.
\]
Its determinant is zero, contradicting the nonvanishing
of the $2\times2$ minors of $A$. Therefore, we have $h\mid s$ or $h\mid s+1$.
Since $2s=q-1$ and $2(s+1)=q+1$, we conclude that
\[
2h\mid q-1~~\text{or}~~2h\mid q+1.
\]
This completes the proof.
\end{proof}

To handle length $2\ell^2$, we also need to pass to a shorter cyclic MDS code. The following lemma uses the zero residue class of the Fourier support for this reduction.

\begin{lem}
	\label{lem:section}
	Let $D$ be the Fourier support of a nontrivial $[n,k]_q$ cyclic MDS code $\cc$.
	For $h\mid n$, put $N=n/h$ and $A=\{a\in\Zz_N:ha\in D\}$ with $c=|A|$.
	Then $A$ is the Fourier support of a cyclic code $\mathcal B$ over
	$\Ff_q$ of length $N$ and dimension $c\in\{\floor{k/h},\ceil{k/h}\}$.
	If $c>0$, then $\mathcal B$ is MDS. If $c=0$, it is the zero code.
\end{lem}
\begin{proof}
Let $m={\rm ord}_n(q)$	and let $\E=\Ff_{q^m}$. Let $\beta$ be a primitive $n$-th root of unity in $\E$. Since $qD=D$, we have $qA=A$ and therefore defines a $q$-ary cyclic code $\mathcal B$ with Fourier support $A$. The length and dimension of $\mathcal B$ is $N$ and $c$, respectively. With $\theta=\beta^h$, the lifted code $\mathcal B_{\E}$ of $\mathcal B$ is
\[
\Span_{\E}\{(1,\theta^a,\ldots,\theta^{(N-1)a}):a\in A\}
\]	
Since $D$ be the Fourier support of a nontrivial $[n,k]_q$ cyclic MDS code, it follows from Lemma \ref{lem:balance} that $c\in\{\floor{k/h},\ceil{k/h}\}$ as $A$ is the zero residue class modulo $h$ of $D$.

Let $\bv_d$ be the vector over $\E$ defined by \eqref{vector-van} and let $\cc_{\E}=\Span_{\E}\{\bv_{a}:a\in D\}$ be the lifted code of $\cc$. For $0\le j<N$ and $0\le b<h$, we have
\[
(\bv_{ha})_{j+bN}=\beta^{ha(j+bN)}=\theta^{aj}.
\]
Then the subcode $W=\Span_{\E}\{\bv_{ha}:a\in A\}$ of $\cc_{\E}$ consists exactly of $h$ consecutive repetitions of codewords of $\mathcal B_{\E}$.
	If $c>0$, we derive that 
	\[
		h\,d(\mathcal B_{\E})=d(W)\ge n-k+1.
	\]
This implies that $d(\mathcal B_{\E})\ge N-\ceil{k/h}+1$.
If $c=\ceil{k/h}$, we deduce that $\mathcal B_{\E}$ and $\mathcal B$ are MDS.
It remains to consider $0<c<\ceil{k/h}$. Then $h\nmid k$ and
	$c=\floor{k/h}$. The corresponding zero residue class of the Fourier support
	of $\cc^\perp$ is
	\[
	\{a\in\Zz_N:ha\in\Zz_n\setminus(-D)\}
	=\Zz_N\setminus(-A).
	\]
	This is precisely the Fourier support of $\mathcal B^\perp$.
	Its size is
	\[
	N-c=N-\floor{k/h}=\ceil{(n-k)/h}.
	\]
	It is thus the larger possible residue-class size for the MDS code
	$\cc^\perp$. Using the similar argument as shown above, we obtain that $\mathcal B^\perp$ is MDS. By Lemma \ref{lem:mds}, $\mathcal B$ is MDS as well.
\end{proof}

\begin{lem}\label{lem:twice}
	Let $r$ and $\ell=2r+1$ be odd primes. Then every nontrivial $[n=2\ell,k]_q$ cyclic MDS
	code satisfies $n\le q+1$
\end{lem}
\begin{proof}
According to Lemma \ref{lem:quadratic}, we need to prove that $2\ell\mid q^2-1$.

Suppose by contradiction that $2\ell\nmid q^2-1$. Since $q$ is odd, it follows from Lemma \ref{lem:fixed} that $\gcd(2\ell,q^2-1)=2$.
	Put $m=\ord_{2\ell}(q)$. Lemma~\ref{lem:boundary} gives
\begin{equation}\label{lem-25-eq-1}
	k\ge2m+1~~\mbox{and}~~2\ell-k\ge2m+1.
\end{equation}
	By Lemma~\ref{cor:divisor}, $m=\ord_\ell(q)\mid2r$.
	Since $m>2$ by our assumption $2\ell\nmid q^2-1$, we have $m\in\{r,2r\}$ and therefore $m\ge r$.
	It follows from \eqref{lem-25-eq-1} that $k\ge2r+1=\ell$ and $2\ell-k\ge\ell$.
	These two inequalities imply that $k=\ell$.
	Lemma~\ref{lem:half} now gives $2\ell\mid q^2-1$, a contradiction. Therefore, we have $2\ell\mid q^2-1$. This completes the proof.
\end{proof}

\begin{lem}\label{lem:twicesquare}
	Let $r$ and $\ell=2r+1$ be odd primes. Let $\cc$ be a nontrivial $[n=2\ell^2,k]_q$ cyclic MDS
	code. Then we have $n\le q+1$.
\end{lem}
\begin{proof}
	Suppose by contradiction that $n>q+1$. By Lemma~\ref{lem:square},
	$q=p^2$ such that $p$ is odd and $\ord_{n}(q)=r$. Since $\ell>3$, it follows from Lemma \ref{cor:divisor} that $\ord_{n}(q)=\ord_{2\ell}(q)=r$.
	According to Lemma~\ref{MDS-2q-2}, we get $2\ell^2\le2p^2-2$, which implies that $p>\ell$.
	
	Replace $\cc$ by its dual if necessary, so that its dimension is
	$\kappa=\min\{k,n-k\}\le\ell^2$ and denote its Fourier support by
	$D$. Since $\gcd(n,q)=1$ and $\ord_n(q)=r>2$, we deduce that $q$ is odd and $q^2\not\equiv 1\pmod n$. Since $\gcd(n,q^2-1)$ is even, it follows from Lemma~\ref{lem:fixed} that $\gcd(n,q^2-1)=2$.
	By the proof of Lemma~\ref{lem:boundary}, exactly one of $0,\ell^2$ belongs to $D$.
	Both indices are divisible by $\ell$.
	
	Using Lemma~\ref{lem:section} with $h=\ell$, we obtain a $[2\ell,c]_q$ cyclic MDS code $\mathcal B$ with $c\in\{\floor{\kappa/\ell},\ceil{\kappa/\ell}\}$. Since exactly one of $0,\ell^2$ belongs to $D$ and $\kappa\leq \ell^2$, we deduce that $1\le c\le\ell$. 
	If $c\ge2$, then $2\le c\le\ell\le2\ell-2$ and therefore
	$\mathcal B$ is nontrivial. Consequently,
	Lemma~\ref{lem:twice} gives $2\ell\mid q^2-1$.
	This contradicts $\ord_{2\ell}(q)=r>2$. Hence $c=1$. Due to $c\in\{\floor{\kappa/\ell},\ceil{\kappa/\ell}\}$, we derive that $2\le\kappa\le2\ell-1$.
	
	The distinct odd primes $p$ and $\ell$ with $p>\ell$ satisfy $p\ge\ell+2$ and hence
	\[
	2\le\kappa\le2\ell-1<2p-2.
	\]
	Lemma~\ref{lem:BD}, applied over $\Ff_{p^2}$ to $\cc$ or its dual of dimension
	$\kappa$, gives $n\le p^2+1=q+1$, a contradiction. This completes the proof.
\end{proof}

For exponents at least three, the reduction to a quadratic extension and the elementary length bound suffice under the stated condition on the characteristic.

\begin{lem}
	\label{lem:high}
	Let $\ell\ge11$ and $r=(\ell-1)/2$ be primes. Let $q=p^f$ with $f$ being a positive integer. Let $e\ge3$ and
	$\varepsilon\in\{1,2\}$. If
	$\ell\ge\ceil{p/2}$, then every nontrivial cyclic MDS code of length
	$n=\varepsilon\ell^e$ satisfies $n\le q+1$.
\end{lem}
\begin{proof}
	Suppose by contradiction that $n>q+1$. Lemma~\ref{lem:square} gives that $p$ is an odd prime and $q=p^2$. Since $\ell$ is an odd prime and $\ell\ge\ceil{p/2}$, we have $p<2\ell$. By $e\ge3$ and Lemma~\ref{MDS-2q-2},
	\[
	\ell^3\le\ell^e\le n\le2p^2-2<8\ell^2.
	\]
	Dividing by $\ell^2$ gives $\ell<8$, contrary to $\ell\ge11$. Hence, we have $n\le q+1$.
\end{proof}

\subsection{Proof of Theorem~\ref{thm:B}}
	We treat the three cases separately.
	
	\medskip\noindent\textit{Case (1).}
	Suppose by contradiction that $n=2^\alpha3^\beta5^\gamma7^\delta>q+1$.
	By Theorem~\ref{cor:A}, $3\nmid n$ and $4\nmid n$. This implies that $\beta=0$ and
	$\alpha\le1$. The same theorem excludes the simultaneous occurrence
	of five and seven, because $\gcd(5-1,7-1)=2$.
	At least one of five and seven must occur, since otherwise
	$n\le2$, contrary to nontriviality.
	Thus $n=\varepsilon5^e$ or $n=\varepsilon7^e$ with
	$\varepsilon\in\{1,2\}$ and $e\ge1$.
	Lemmas~\ref{lem:five} and~\ref{lem:seven} exclude both possibilities. Hence, we have $n=2^\alpha3^\beta5^\gamma7^\delta\leq q+1$
	
	\medskip\noindent\textit{Case (2).}
	Suppose by contradiction that the length $n$ in Case (2) satisfies $n>q+1$.
	Theorem~\ref{cor:A} again gives $\alpha=1$ and $\beta=0$.
	For each $i=1,\cdots,s$, write $\ell_i=2r_i+1$, where $r_i\ge5$ is prime.
	Let $A=\{5,7,\ell_1,\ldots,\ell_s\}$. For any two distinct elements $x,y\in A$, it is clear that $\gcd(x-1,y-1)=2$.
	By Theorem~\ref{cor:A} and $n>2$, exactly one odd prime $s\in A$ can divide $n$.
	
	If $s=5$ or $s=7$, then $n=2\cdot5^e$ or
	$n=2\cdot7^e$. By Lemmas~\ref{lem:five} and~\ref{lem:seven}, we obtain $n\leq q+1$.
	Otherwise $n=2\ell_i^{e_i}$ for a unique $i$ with $e_i\ge1$.
	For $e_i=1$ or $2$, Lemmas~\ref{lem:twice} and~\ref{lem:twicesquare} give $n\leq q+1$. 
	For $e_i\ge3$ with $\ell_i\ge\ceil{p/2}$, it follows from Lemma~\ref{lem:high} that $n\leq q+1$.  
	Therefore, every possibility contradicts $n>q+1$. 
	
	\medskip\noindent\textit{Case (3).}
	Here $n=\ell_i^e$, $e\ge3$, and $\ell_i\ge\ceil{p/2}$.
	Apply Lemma~\ref{lem:high} with $\varepsilon=1$.
	This gives $n\le q+1$ and completes the proof.

\section{Conclusion}\label{sec:conclusion}

For a nontrivial simple-root $[n,k]_q$ cyclic MDS code, we proved $n\le q+1$ when $\gcd(n,q^2-1)>2$, or when $k$ is even and $\gcd(n,q^2-1)>1$. The bound also holds when $3\mid n$ or $4\mid n$, and when $n$ has distinct odd prime divisors $\ell_1,\ell_2$ with $\gcd(\ell_1-1,\ell_2-1)=2$. We established the same bound for all lengths whose prime factors lie in $\{2,3,5,7\}$. Further cases involve primes $\ell\ge11$ for which $(\ell-1)/2$ is prime, under the conditions in Theorem~\ref{thm:B}.

The proofs combine Fourier supports, locality bounds, multiplicative orders, field descent, and valuation estimates. The arithmetic restrictions used here leave many cyclic lengths untreated, and extending the results to these lengths remains a natural problem. The full MDS conjecture also includes codes without cyclic symmetry. We expect that new techniques will be needed to address the remaining cases and ultimately resolve the conjecture.

\section*{Acknowledgements}

The authors used OpenAI Codex (GPT-6), with the Mathbox plugin, to assist with manuscript revision under their direction. This assistance included editing the language and exposition, checking local notation and proof consistency. 

%

\providecommand{\bysame}{\leavevmode\hbox to3em{\hrulefill}\thinspace}
\providecommand{\MR}{\relax\ifhmode\unskip\space\fi MR }
\providecommand{\MRhref}[2]{%
	\href{http://www.ams.org/mathscinet-getitem?mr=#1}{#2}
}
\providecommand{\href}[2]{#2}

\end{document}